\documentclass[aps,pre,preprint,groupedaddress,showpacs]{revtex4}
\usepackage{graphicx}

\begin{document}

\title{Intermittency and rough-pipe turbulence}

\author{Mohammad Mehrafarin and Nima Pourtolami }
\affiliation{Physics Department, Amirkabir University of
Technology, Tehran 15914, Iran} 
\email{mehrafar@aut.ac.ir}

\begin{abstract}
Recently, by analyzing the measurement data of Nikuradze, it has been proposed (N. Goldenfeld, Phys. Rev. Lett.
{\bf{96}}, 044503, 2006) that the friction factor, $f$, of rough pipe flow obeys a scaling law in the turbulent regime. Here, we provide a phenomenological scaling argument to explain this law and demonstrate how intermittency modifies the scaling form, thereby relating $f$ to the intermittency exponent, $\eta$. By statistically analyzing the measurement data of $f$, we infer a satisfactory estimate for $\eta$ ($\approx 0.02$), the inclusion of which is shown to improve the data-collapse curve. This provides empirical evidence for intermittency other than the direct measurement of velocity fluctuations. 
\end{abstract}


\maketitle

\section{Introduction}
A major aspect of fully developed turbulence is the existence of universal scaling laws in the so called inertial-range scales, $l_0\ll l\ll l_d$, $l_0$ 
($l_d$) being the length scale of the energetic (dissipative)-range eddies. In particular, the scaling behavior of the 2nd-order velocity structure function under the Kolmogorov refined similarity hypothesis \cite{Kolmogorov} is expressed as
\begin{equation}
\langle \delta v_l^2 \rangle \sim (\langle \varepsilon_l \rangle l)^{\frac{2}{3}} \sim l^{\frac{2}{3}+\eta} \label{1}
\end{equation}
where $\langle \varepsilon_l \rangle$ is the average dissipation rate over a sphere of size $l$ and $\eta$ is the intermittency exponent, whose value is considered universal.

An important physical quantity in rough pipe flow is the friction factor $f$, which is related to the pressure drop across the pipe according to the Darcy-Weisbach formula (see e.g. \cite {Shames}). The friction factor is a function of the Reynolds number, $Re$, of the flow and the relative roughness, $\frac{r}{R}$, of the pipe ($r$ being the average size of the roughness elements and, $R$, the radius of the pipe). $f$ can be expressed in terms of the wall stress,
$\tau$, as $f=\frac{\tau}{\rho V^{2}}$, where $\rho$ is the density
and $V$ is the cross-sectional average of the mean-time-average velocity of the flow.

In a seminal series of experiments on rough pipe turbulence that has remained a benchmark in the field,
Nikuradze \cite
{Nikuradze} elucidated how $f(\frac{r}{R},Re)$ depends on its arguments. His data was presented as six curves, which are shown in Fig. 1. The main features of these plots are as follows. Up to $Re\sim 3300$, the flow is still laminar and $f\sim\frac{1}{Re}$. This, of course, corroborates the exact result $f=\frac{64}{Re}$, obtained from the Navier-Stokes equation. At $Re\sim 3500$, the entire
curves plunge to the ``smooth-pipe zone" in accord with Blasius's
scaling \cite {Blasius,Blasius1}, $f\sim Re^{-\frac{1}{4}}$. Here the friction factor is independent of boundary roughness, due to the existence of a sufficiently thick viscous sublayer near the wall that screens the roughness elements from the turbulent flow. As $Re$ increases, the roughness elements become progressively exposed to the turbulent flow and at large enough $Re$, the flow enters the ``rough-pipe zone", where the friction factor becomes dependent only on boundary roughness in accord with Strickler's
law \cite{Strickler}, $f\sim (\frac{r}{R})^{\frac{1}{3}}$.

\begin{figure} 
\includegraphics[width=10cm]{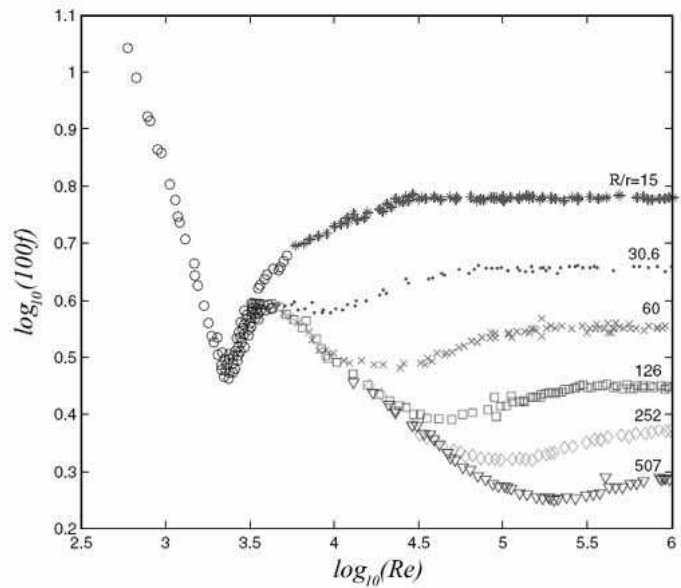}
\caption{Friction factor, $f$, of rough pipe flow versus
Reynols number, $Re$, for different values of relative roughness, $\frac{r}{R}$, in Nikuradze's experiment.}
\end{figure}

These general characteristics can also be seen in other pipe
\cite{Colebrook1,Colebrook2,Sletfjerding} and open-channel
\cite{Shames,Chow} flow data.

Recently, Goldenfeld has pointed out \cite{Goldenfeld} that
Nikuradze's data conform to the scaling form
\begin{equation}
f=Re^{-\frac{1}{4}}\ g(\frac{r}{R}\ Re^{\frac{3}{4}}) \label{2}
\end{equation}
where the scaling function $g(x)$ has the asymptotic behavior,
$$
g(x) \sim \left\{\begin{array}{rl} 
$const.$, &\mbox{$x \rightarrow 0$}\\
x^{\frac{1}{3}},\ \ \ &\mbox{$x \rightarrow \infty$} \end{array} \right. 
$$
and data collapse for $fRe^{\frac{1}{4}}$ versus $\frac{r}{R} Re^{\frac{3}{4}}$ occurs for data that lie between the Blasius and Strickler regimes. The residual deviation of data collapse from the scaling form (\ref{2}), apart from uncertainties in the data, may also reflect something more fundamental \cite{Goldenfeld}. Here, we address the question as to what extent (if any) intermittency corrections, which have not been included in the original analysis, improve the data-collapse curve.

Although turbulent flow in a pipe is anisotropic and inhomogeneous, Kolmogorov's theory, which rest's on the assumptions of isotropy and homogeneity, still applies
\cite{Bombardelli,Gioia,Knight,Lundgren,Lundgren2}. Here, using (\ref{1}), we provide a scaling argument to explain the scaling form (\ref{2}) and demonstrate how it is modified by intermittency, thereby relating $f$ to $\eta$. The measurement data of $f$ can be, thence, utilized to infer the numerical value of $\eta$, the inclusion of which is shown to improve the data-collapse curve. By statistically analyzing Nikuradze's data, we thus obtain an estimate for $\eta$ ($\approx 0.02$) that reconciles with the available values \cite{Kolmogorov,Benzi,Benzi2,She,Chevillard,Frisch}. This provides empirical evidence for intermittency other than the direct measurement of velocity fluctuations.

\section{Phenomenological scaling argument}
The shear stress exerted by the flow on the wall of the pipe scales as \cite{Bombardelli,Gioia} $\tau\sim\rho Vv_r$, where
$v_r=\sqrt{\langle\delta v_r^2\rangle}$ is the velocity of eddies of length scale r, the size of roughness elements. (Such eddies have dominant role in the momentum transfer to the wall
\cite{Bombardelli,Gioia}). Thus, the friction factor can be written as
$f\sim \frac{v_r}{V}$ and hence from (\ref{1}), $f\sim r^\alpha$ where $\alpha=\frac{1}{3}+\frac{\eta}{2}$. Let us now rescale such that $r\rightarrow lr$. Under this rescaling, we have $f\rightarrow l^\alpha f$. Moreover, for $l>1$ ($l<1$), the rescaling results in a mere amplification (de-amplification) of roughness elements, without changing their geometry, thereby rendering the roughness to be exposed to the turbulent flow at lower (higher) $Re$. Thus, under the rescaling $r\rightarrow lr$, we expect the
same flow, albeit with $Re\rightarrow l^{-\beta}Re$, where
$\beta>0$. That is, under rescaling,
$$
f(\frac{r}{R},Re)\rightarrow f(l\frac{r}{R},l^{-\beta}Re)
$$
Collecting results, we therefore have
\begin{equation}
f(l\frac{r}{R},l^{-\beta}Re)=l^\alpha f(\frac{r}{R},Re) \label{3}
\end{equation}

Equation (\ref{3}) indicates that $f$ is a generalized homogeneous function of its arguments in the inertial range and the behavior is self-affine. We can expound further on the exponent $\beta$, if we bear in mind that we are considering boundary layer turbulence. The turbulent transfer of momentum to the boundary wall by eddies, which predominantly takes place on length scale $r$ (the size of dominant eddies), gives rise to an extra (eddy) viscosity near the wall.
The effective viscosity in the boundary layer, therefore, scales as $\sim rv_r$ (length $\times$ velocity), i.e. as $\sim r^{\alpha+1}$. Hence on rescaling, $Re\rightarrow l^{-\alpha-1}Re$,
i.e. $\beta=\alpha+1=\frac{4}{3}+\frac{\eta}{2}$ and (\ref{3}) reads:
\begin{equation}\label{4}
f(l\frac{r}{R},l^{-\alpha-1}Re)=l^{-\alpha}f(\frac{r}{R},Re)
\end{equation}
Goldenfeld's scaling form (\ref{2}) is just a
particular form of (\ref{4}), which can be obtained by taking
$l=Re^{\frac{1}{\alpha+1}}$ and $\eta=0$, of course. In presence of intermittency, the scaling form thus becomes
\begin{equation}
f=Re^{-\frac{2+3\eta}{8+3\eta}}\ g(\frac{r}{R}\ Re^{\frac{6}{8+3\eta}})\label{5}
\end{equation}
where, asymptotically,
$$
g(x) \sim \left\{\begin{array}{rl} 
$const.$, &\mbox{$x \rightarrow 0$}\\
x^{\frac{1}{3}+\frac{\eta}{2}},  &\mbox{$x \rightarrow \infty$} \end{array} \right. 
$$
Note that the Blasius and Strickler laws are, therefore, modified according to
$$
f\sim Re^{-\frac{2+3\eta}{8+3\eta}},\ \ \ \
f\sim \left( \frac{r}{R} \right)^{\frac{1}{3}+\frac{\eta}{2}}
$$
respectively. The above modified Strickler formula coincides (to lowest order) with the result obtained by Gioia {\it et. al.} \cite{Bombardelli}. The modified Blasius formula, however, does not. The source of discrepancy lies in their application of the expression $Re^{-\frac{3}{4}}$ for the Kolmogorov scale, which is invalid when the intermittency exponent is non-zero. The correct expression is $Re^{-\frac{6}{8+3\eta}}$, the application of which would have had resulted in the same modified Blasius formula as given above \cite{Goldenfeld2}.

Scaling form (\ref{5}) relates the friction factor to the intermittency exponent and provides ground for the estimation of the latter via empirical data other than the direct measurement of velocity fluctuations. 

\section{Estimation of the intermittency exponent}

Having established (\ref{5}), we proceed to infer the value of $\eta$ from the measurement data of $f$. To this end, we plot $fRe^{\frac{2+3\eta}{8+3\eta}}$ versus
$\frac{r}{R}Re^{\frac{6}{8+3\eta}}$ for different trial values of $\eta$ and examine the resulting data collapse by applying regression analysis (see e.g. \cite{Netter}). The best value for $\eta$ is one that will result in the highest possible data correlation, i.e., the smallest possible data scatter (deviations from data collapse curve). The data correlation coefficient, $\rho$, is defined by the fraction of the total variation (sum of squares) of data, $y$, that is explained by the regression model (fitting curve), according to
$$
\rho^{2}=\frac{\sum(\hat{y}-\bar{y})^2}{\sum(y-\bar{y})^2}.
$$
Here $\bar{y}$ represents the average of data values, $y$, and $\hat{y}$ are their estimated values based on the regression curve. The closer $\rho$ to 1, the more correlated the data, and the better the regression is. For a given regression model, we therefore examine the variation of $\rho$ with
$\eta$; the best estimate of $\eta$ is the one that yields the correlation coefficient closest to 1. Now polynomial regression is suitable for nonlinear data curves that show a maximum/minimum and the absence of this feature in the data collapse curve, as shown in Fig. 2 of reference
\cite{Goldenfeld} (a non zero value of $\eta$ does not change this feature), renders the polynomial model irrelevant.
We, therefore, resort to the exponential regression model, the result of which is presented in
Fig. 2. The estimated value of $\eta$ thus obtained is 0.02, which agrees with the generally expected range of values ($\approx 0.02 - 0.03$) reported in the literature \cite{Kolmogorov,Benzi,Benzi2,She,Chevillard,Frisch}. That is to say, Nikuradze's data set best conforms to $\eta \approx 0.02$ and not, in particular, to $\eta=0$. It is rather interesting that large-scale properties like friction factors can provide evidence for intermittency, which is a direct manifestation of the small-scale statistics. This and similar previous observations \cite{Grossmann} embody deep connections between spectral structure and the global properties of turbulent systems.
\begin{acknowledgments}
We thank Dr. N. Goldenfeld for bringing to our attention the discrepancy concerning the modified Blasius formula.
\end{acknowledgments}

\newpage
\begin{figure}
\includegraphics[width=10cm]{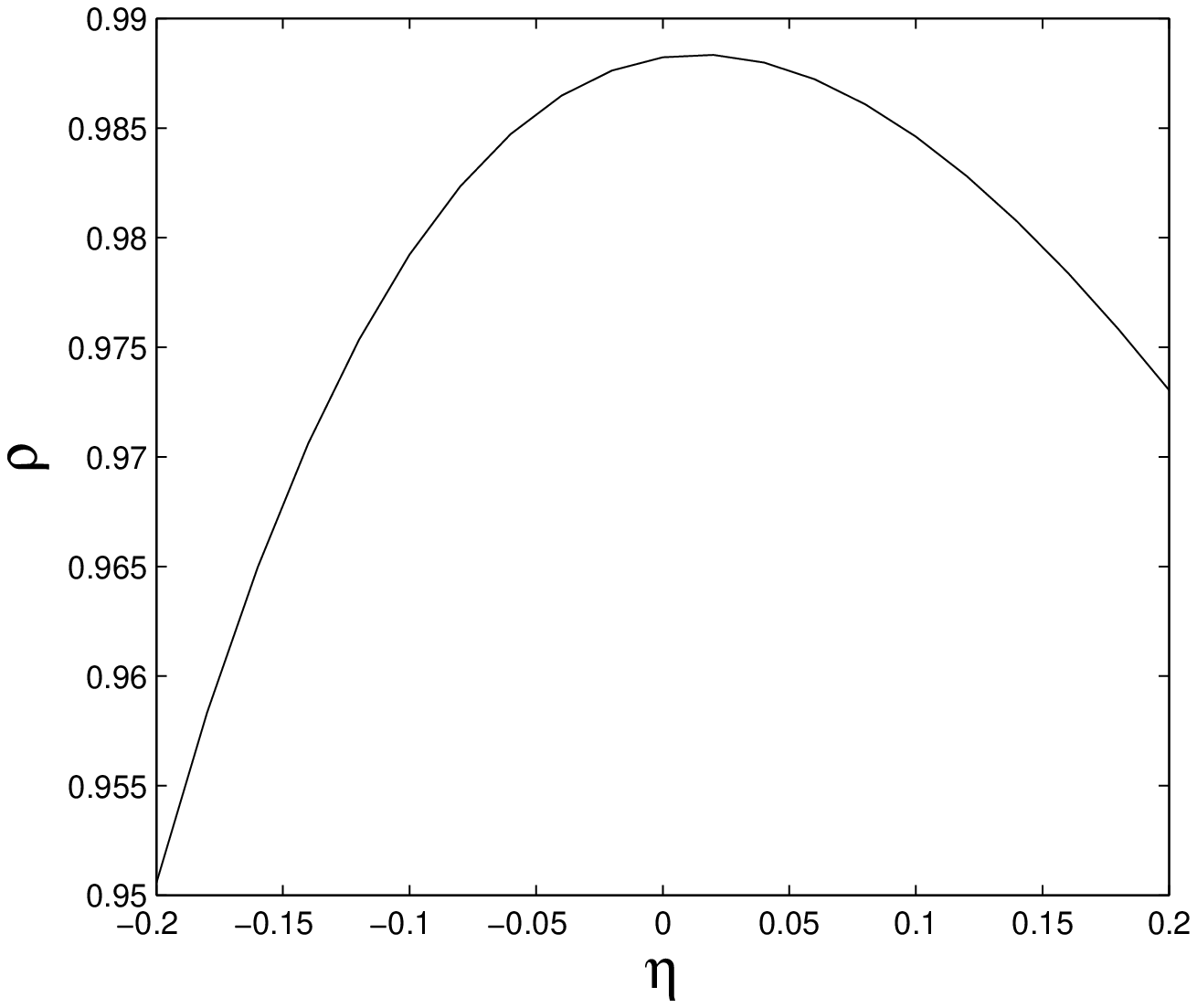}\\
\includegraphics[width=10cm]{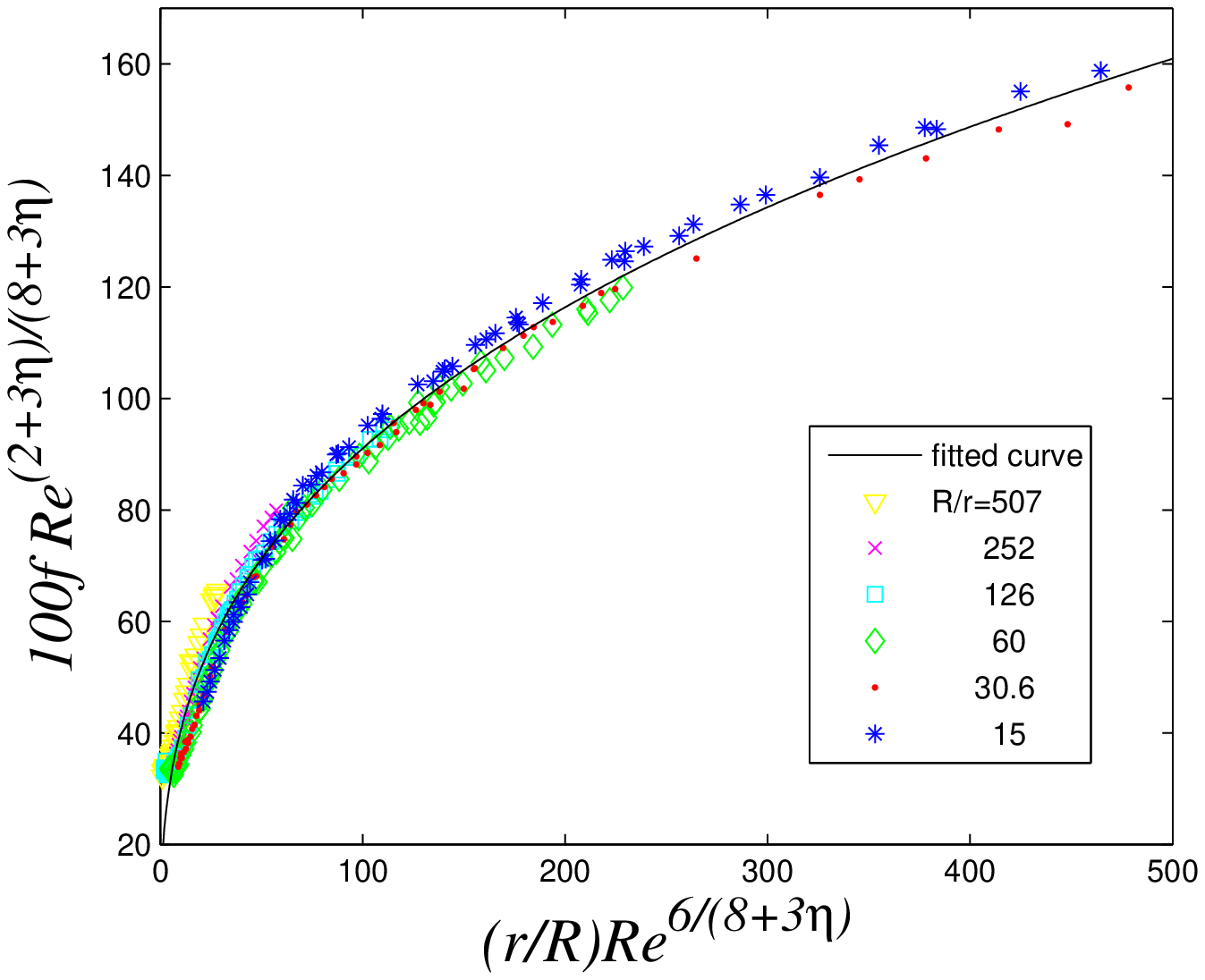}\\
\caption{(Color online.) Exponential regression model. Best data correlation is obtained for the value $\eta=0.02$ (top), resulting in the best data collapse (bottom).}
\end{figure}


\begin{thebibliography}{widest-label}

\bibitem{Kolmogorov} A. N. Kolmogorov, J. Fluid Mech. {\bf 13}, 82 (1962).
\bibitem{Shames} I. H. Shames, {\it Mechanics of fluids}, 4th ed. (McGraw-Hill, New York, 2003).
\bibitem{Nikuradze} Reprinted in english in J. Nikuradze, NACA Tech. Memo. {\bf 1292}, (1950).
\bibitem{Blasius} H. Blasius, Forsch. Arb. Ing. Wes. (Berlin) No. 134 (1913).
\bibitem{Blasius1} H. Schlichting, {\it Boundary layer theory} (McGraw-Hill, New York, 1979).
\bibitem{Strickler} Reprinted in English in A. Strickler, {\it Contributions to the question of a velocity formula and roughness data for streams, channels and
closed pipelines}, translation by T. Roesgan and W. R. Brownie,
CIT, Pasadena (1981).
\bibitem{Colebrook1} C. F. Colebrook and C. M. White, Proc. R. Soc. A {\bf 161}, 367 (1937).
\bibitem{Colebrook2}  C. F. Colebrook, Inst. Civ. Eng. J. {\bf 11}, 133 (1939).
\bibitem{Sletfjerding}  E. Sletfjerding and J. S. Gudmundsson, J. Energy Resour. Technol. {\bf 125}, 126 (2003).
\bibitem{Chow}  V. T. Chow, {\it Open-channel hydraulics} (McGraw-Hill, New York, 1988).
\bibitem{Goldenfeld}  N. Goldenfeld, Phys. Rev. Lett. {\bf 96}, 044503 (2006).
\bibitem{Bombardelli} G. Gioia, P. Chakraborty and F. A. Bombardelli, Phys. Fluids {\bf 18}, 038107 (2006).
\bibitem{Gioia} G. Gioia and P. Chakraborty, Phys. Rev. Lett. {\bf 96}, 044502 (2006).
\bibitem{Knight}  B. Knight and L. Sirovich, Phys. Rev. Lett. {\bf 65}, 1356 (1990).
\bibitem{Lundgren} T. S. Lundgren, Phys. Fluids {\bf 14}, 638 (2002).
\bibitem{Lundgren2} T. S. Lundgren, Phys. Fluids {\bf 15}, 1074 (2003).
\bibitem{Benzi} R. Benzi, S. Ciliberto, R. Tripiccione, C. Baudet, F. Massaioli and S. Succi, Phys. Rev. E {\bf 48}, R29 (1993).
\bibitem{Benzi2} R. Benzi, S. Ciliberto, C. Baudet and G. Ruiz Chavarria, Physica D {\bf 80}, 385 (1995). 
\bibitem{She} Z. S. She and E. Leveque, Phys. Rev. Lett. {\bf 72}, 336 (1994).
\bibitem{Chevillard} L. Chevillard, B. Castaing, E. Leveque and A. Arneodo, Physica D {\bf 218}, 77 (2006). 
\bibitem{Frisch} see also U. Frisch, {\it Turbulence: The legacy of A. N. Kolmogorov} (Cambridge University Press, Cambridge, 1995), Ch. 8. \bibitem{Goldenfeld2} N. Goldenfeld, private communication.
\bibitem{Netter} J. Netter, W. Wasserman and G. A. Whitmore, {\it Applied statistics}, 3rd ed. (Allyn and Bacon, Massachusetts, 1988). 
\bibitem{Grossmann} S. Grossmann, Phys. Rev. E {\bf 51}, 6275 (1995).

\end{thebibliography}
\end{document}